\documentclass[prl,aps,epsfig,twocolumn,showpacs]{revtex4}
\usepackage{graphicx}
\begin{document}

\title{Generalized scaling relations for unidirectionally coupled 
nonequilibrium systems}

\author{Sungchul Kwon and Gunter M. Sch\"utz}
\affiliation{
Institut f\"ur Festk\"orperforschung, Forschungszentrum J\"ulich,
D-52425 J\"ulich, Germany
}
\date{\today}

\begin{abstract}
Unidirectionally coupled systems which exhibit phase transitions 
into an absorbing state are investigated at the multicritical point.
We find that
for initial conditions with isolated particles, each hierarchy level
exhibits an inhomogeneous active region, coupled and uncoupled 
respectively. The particle number of each level increases algebraically 
in time as $N(t) \sim t^{\eta}$ with different exponents
$\eta$ in each domain.
This inhomogeneity is a quite general feature of unidirectionally
coupled systems and leads to two hyperscaling relations between 
dynamic and static critical exponents.
Using the contact process and the branching-annihilating random walk with 
two offsprings, which belong to the DP and PC 
classes respectively, we numerically confirm the scaling relations.   

\end{abstract}

\maketitle

Nonequilibrium absorbing phase transitions (APT) have been studied 
extensively over the last decade in the hope of understanding better 
nonequilibrium continuous phase transitions \cite{review,hin}. 
Also APT are one of the simplest and natural extensions of the
well-established equilibrium transitions to nonequilibrium systems.
In APT, a system undergoes a continuous phase transition from an 
active into an absorbing state where no dynamics occur furthermore. 
Models exhibiting APT describe wide range of phenomena such as 
epidemic spreading, catalytic chemical reactions, surface growth, 
wetting and roughening transitions, self-organized criticality, and 
transport in disordered media \cite{review,hin}.
 
The basis of theoretical and numerical analysis is the understanding 
that the concept of universality plays a central role in classifying
nonequilibrium phase transitions as it does in thermal equilibrium. 
However, surprisingly few universality classes for APT
have been identified so far, and  -- as a unifying theoretical framework 
is not available -- we are still far from a systematic classification of 
APT even in one spatial dimension. The best-studied directed percolation (DP) 
class includes the systems which have no symmetry between absorbing states 
and no conservation laws in the order parameter \cite{review,hin}.
Another well-known class is the parity-conserving (PC) class which 
describes models with $Z_2$ symmetry between absorbing states or parity 
conservation of the number of order parameter \cite{hin,pc}. 
Recent studies show that the conservation of total number of
particles and coupling with auxiliary diffusion fields also form 
distinct classes \cite{cons,pcpd}, or give rise to unexpected new
critical exponents \cite{critex}.

Several critical exponents characterize the off-critical and critical behavior 
of APT. The exponents $\beta$, $\nu_\bot$, $\nu_{||}$ characterize the scaling 
behavior of
steady-state particle density $\rho^s$, the correlation length $\xi$,
characteristic time $\tau$ near criticality in thermodynamic limit as \cite{static}
\begin{equation}
\rho^s \sim \Delta^{\beta},\; \xi \sim |\Delta|^{- \nu_{\bot}} ,\; 
\tau \sim |\Delta|^{-\nu_{||}},
\end{equation}
where $\Delta$ is the distance from criticality.
At criticality, $\delta$, $\alpha$, $\eta$ and $z$ characterize the dynamic scaling of
survival probability of a single particle $P_s (t)$, 
density of particles $\rho(t)$,
the number of particle $N(t)$ averaged over all samples 
and spreading distance $R(t)$ averaged over survival samples as \cite{Grass}
\begin{equation}
P_s (t) \sim t^{-\delta},\; \rho (t) \sim t^{-\alpha},\; N(t) \sim t^\eta , 
\; R(t) \sim t^{1/z} ,
\end{equation}
with $\alpha = \beta/\nu_{||}$, $z = \nu_{||}/\nu_{\bot}$ and $d$ is 
the spatial dimension.
The well-known scaling relation between these exponents is \cite{Grass}
\begin{equation}
\eta = d/z - \alpha - \delta \; .
\end{equation}
Except in systems with infinitely many absorbing states, 
we have $\alpha = \delta$ \cite{ima}.

Rather than exploring new systems exhibiting unknown critical behavior, 
coupled systems of known universality classes have been studied recently
as an another direction of searching for new universality classes.
A coupled systems is a multi-species system in which each species is coupled to 
the others in certain ways. Among possible ways of coupling such as bidirectional,
cyclic and unidirectional coupling in linear or quadratic way \cite{uni,coupling,nbaw}, 
a recent study on a hierarchy of linearly unidirectionally coupled DP systems 
shows that a new critical behavior emerges at the multicritical point 
where criticality of all levels coincide \cite{uni}. 
The multicritical behavior is characterized by a varying
order parameter exponent $\beta$, taking
a different value in each level of the hierarchy, while the
exponents $\nu_\bot$ and $\nu_{||}$ are those of the DP class in all levels. 
Also $\eta$ and $\delta$ vary according to the levels of the hierarchy,
but these exponents do not satisfy the ordinary scaling relation of Eq. (3)
in each higher level $k(>1)$.
Instead they are suggested to satisfy \cite{uni}
\begin{equation}
\eta_k = d/z_k - \alpha_k -\delta_1
\end{equation}
with $\alpha_k = \delta_k$.
In Eq. (4), the exponent of $P_s (t)$ of the first level, $\delta_1$
replaces $\delta_k$ of $k$' level. The appearance of $\delta_1$ in Eq. (4) 
is unusual and there has been no clear understanding for that.
It should be noted that Eq. (4) comes from taking an average
where one uses configurations in which 
each level survives regardless of the survival of the first level. 
We shall refer to this usual average as $slave$ average to distinguish it from our
average method discussed below. 

In this paper, we report generalized scaling relations for linearly and 
unidirectionally coupled systems belonging to same or different universality
classes. 
We propose an explanation for the appearance of $\delta_1$ in Eq. (4).
We also show that Eq. (4) is not satisfied in general coupling systems because
it is a consequence of the usual slave average which does not take into
account an important inhomogeneity in the distribution of particles
in the higher levels of the hierarchy.
In addition,  our scaling relations provides insight in
the origin of markedly large corrections to scaling
which make estimations of critical exponents difficult.

We consider a hierarchy of an arbitrary number of coupled systems
with coupling $A \rightarrow B \rightarrow C \cdots $.
For the derivation of scaling relations, 
we are interested in the dynamic behavior of the $k$'th level starting 
with a single particle on the first level. Then on each level, a cluster is 
created and spreads. Due to the coupling, the size of the cluster on each 
level is always larger than those of the lower levels.
It means that the cluster of the $k$'th level (with size $R_k (t)$) is 
divided into two parts, the coupled and the uncoupled region where the
dynamics on level $k$ evolve autonomously (Fig. 1).
In the coupled region (with size $R_C (t)$), the $(k-1)$'th level feeds
particles to the $k$'th level so that $R_C (t)$ is actually the spreading distance 
of the $(k-1)$'th level ($R_{k-1} (t)$). 
However, in the uncoupled region (with size $R_U (t)$), there is only the feeding 
from the boundary of
$R_C (t)$. So in $R_U (t)$, the cluster evolves with its own reaction dynamics.
It implies that the number of particles in the $k$'th level ($N_k (t)$)
increases in time with different exponents at the multicritical point, viz.,
$N_C \sim t^{{\eta^C}_k}$ and $N_U \sim t^{{\eta^U}_k}$ 
in the coupled and uncoupled area
respectively. It is clear that this inhomogeneity of the total particle
number $N_k (t)=N_C (t)+N_U (t)$ is a general property
of unidirectionally coupled systems.

The significance of this inhomogeneity is illustrated by noting that
at the multicritical point, each level decays more slowly than its lower levels.
So the coupling to lower levels is always broken after some time.
To exploit the inhomogeneity of $N_k (t)$ for any $k>1$
we only consider the configurations in which the source ($k=1$) still survives. 
In these configurations, slave systems ($k>1$) always survives
with unit probability so the survival probability $P_s (t)$ of the slave is just 
that of the source and
we have $\delta_k \neq \alpha_k$ and $\delta_k = \delta_1$ for any $k>1$.
We call these configurations as $source$ ensemble and the average
with this ensemble as $source$ average. 
As shown below the source average exhibits more clearly the special properties of 
unidirectionally coupled systems than the usual slave average and allows for the
derivation of the scaling 
relations without any assumptions of scaling functions for observables.

\begin{figure}
\includegraphics[scale=0.2]{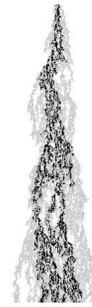}
\caption{\label{fig-1}
The evolution of a cluster of the second level of CP-CP coupling. 
Black(gray outside black) corresponds to coupled(uncoupled) region. 
}
\end{figure}

In a coupled region, $N_C (t)$ is given by the relation, 
$N_C (t) = \rho_C (t) \times {R_C}^d (t) \times {P_s}^{k=1}(t) $, where $\rho_C (t)$ and
${P_s}^{k=1}(t)$ are the particle density of $k$'th level inside of the volume 
${R_C}^d (t)$ and the survival probability of the first level.
Assuming the power-law decay of $\rho_C$, $R_C$ and ${P_s}^{k=1}$ as 
$\rho_C \sim t^{-{\alpha^C}_k}$, $R_C = R_{k-1} \sim t^{1/z_{k-1}}$ and 
${P_s}^{k=1} \sim t^{-\delta_1}$,
we obtain the first scaling relation
\begin{equation}
{\eta^C}_k = d/z_{k-1} - {\alpha^C}_k - \delta_1 \; .
\end{equation}
In the uncoupled region, ${\rho^U}_k \sim t^{-{\alpha^U}_k}$ and
$R_U \sim t^{1/{z^U}_k}$. Then similarly we find the
second relation
\begin{equation}
{\eta^U}_k = d/{z^U}_k - {\alpha^U}_k - \delta_1 \; .
\end{equation}
From $R_k = R_U + R_{k-1} \sim t^{1/z_k}$, 
we have a condition, ${z^U}_k \geq z_k $.  
The above two relations are generalized scaling relations for unidirectionally
coupled systems taking the inhomogeneity into account.

Our scaling relations are very different from Eqs. (3) and (4)
because they are not scaling relations for quantities of the whole system.
The number of total particles $N_k (t)$ and total density $\rho_k (t)$ 
scale as
$N_k (t) = N_C (t) + N_U (t) \sim t^{\eta_k}$ and 
$\rho_k (t) = \rho_C (t) + \rho_U (t) \sim t^{-\alpha_k}$. 
Therefore $\eta_k$ should be the larger one of ${\eta^C}_k$ and ${\eta^U}_k$, 
while $\alpha_k$ be the smaller one of ${\alpha^C}_k$ and ${\alpha^U}_k$. 
For ${z^U}_k = z_{k-1}$, ${\eta^C}_k$ is larger than ${\eta^U}_k$ because 
${\alpha^C}_k$ and ${\alpha^U}_k$ always satisfy ${\alpha^C}_k \leq {\alpha^U}_k$. 
In that case, we have $\eta_k = {\eta^C}_k$ and $\alpha_k = {\alpha^C}_k$ and 
we get 
$\eta_k = d/z_k + \alpha_k + \delta_1$ with $\alpha_k \neq \delta_k$. 
This is of the same form as Eq. (4) for ${z^U}_k = z_{k-1}$ 
but with $\alpha_k \neq \delta_k$.
However the values of $\eta_k$ is not the same in general
because it is measured in different ways. 

Notice that the inhomogeneity of a cluster gives corrections,
$t^{-\Delta_\rho}$, $t^{-\Delta_\eta}$ and $t^{-\Delta_z}$ with some negative 
coefficients to the scaling with the asymptotic exponents
$\alpha_k$, $\eta_k$ and $z_k$. The quantities
$\Delta_\rho$, $\Delta_\eta$ and $\Delta_z$ are defined as 
$\Delta_\rho =|{\alpha^C}_k - {\alpha^U}_k|$, 
$\Delta_\eta = |{\eta^C}_k - {\eta^U}_k |$ and
$\Delta_z = d| 1/z{^U}_k -1/z_{k-1}|$ respectively.
These correction exponents may be small so that they can cause a significant 
long time drift of effective exponents measured in simulations.
This makes the precise estimation of exponents very difficult 
in unidirectionally coupled systems.

\begin{figure}
\includegraphics[scale=0.6]{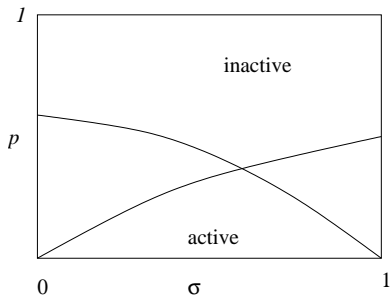}
\caption{\label{fig-2}
A schematic phase diagram of a two-level hierarchy. 
The region labeled by 'active(inactive)' corresponds to an active(inactive)
phase of both levels. In the other regions, each level is in a different phase.
}
\end{figure}

We test the scaling relations Eq. (5) and (6) in a two-level hierarchy 
in one spatial dimension. Each system belongs to the DP or PC class.
We call the first and the second level $A$ and $B$ respectively.
Then for the second level we have the scaling relations
\begin{equation}
\begin{array}{cc}
\eta_C = 1/z_A - \alpha_C - \delta_A \; , \\ 
\eta_U = 1/z_U - \alpha_U - \delta_A \; ,
\end{array}
\end{equation}
where we omit the index $B$ from $\eta$ and $\alpha$.
We denote the coupling of two systems as $Source$-$Slave$ coupling, where 
the source system feeds particles to slave one.

There are four ways of coupling DP and PC system. 
For PC-PC coupling, the existence of a multicritical point depends on the way of
coupling \cite{pcpc,nbaw}. We will discuss it in other place.
For DP-DP coupling, we consider a two-level hierarchy of contact process (CP) \cite{cp}.
In the first level, a particle $A$ is spontaneously annihilated with probability 
$p$ and it creates one $A$ particle on one of the nearest neighboring ($nn$) 
sites with $(1-\sigma)(1-p)$.  
In the second level, the annihilation and branching occur with $p$ and 
$\sigma(1-p)$ respectively.
In branching, if the target site is already occupied by an other
particle, the branching process is rejected.
The probability $\sigma$ is introduced for the coupling.
The criticality of each level is ${p_C}^A = {p_C}^B = 0.232674(4)$ 
at $\sigma = 0$ for the first and $\sigma=1$ for the second level \cite{cp}.
Using the fact that the ratio of creation and annihilation probabilities, 
$R = (1-p_c )/p_c$ should be same at criticality for any value of $\sigma$, 
we find that the critical line of each level in $\sigma-p$ phase diagram is  
${\sigma_c}^{A} = 1- p R/(1-p)$ and ${\sigma_c}^B = p R / (1-p)$ with 
$R = 3.29785$ (Fig. 2).
The coupling dynamics $A \rightarrow A + B $ with $\sigma(1-p)$ 
linearly and unidirectionally couple two CPs without feedback from the
second to the first level. The
multicritical point $(\sigma_M , p_M)$ is the intersection point of the
two critical lines, ${\sigma_c}^A$ and ${\sigma_c}^B$. 
We find $\sigma_M = 1/2$ and $p_M = 0.13165$.
If the multicritical point is approached along the line
$\Delta \equiv (p_M - p)\rightarrow 0$ with fixed $\sigma = \sigma_M $,
new critical behavior emerges \cite{uni}.

\begin{figure}
\includegraphics[scale=0.5]{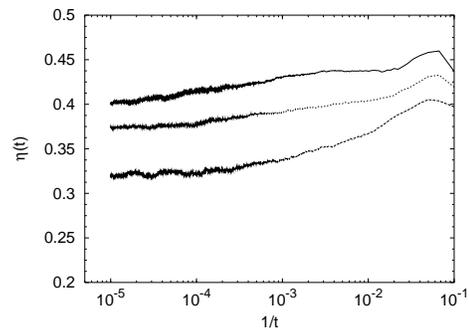}
\caption{\label{fig-3}
Plot of the effective exponent of the second level in CP-CP coupling at 
the multicritical point. The lines correspond to $\eta_C$, $\eta_B$ and $\eta_U$ 
(from top to bottom).
}
\end{figure}

By measuring $\eta_C$ and $\eta_U$ of Eq. (7), we can predict $\eta_B$ and
$\alpha_B$ defined as $N_B = N_C + N_U \sim t^{\eta_B}$ and 
$\rho_B = \rho_C + \rho_U \sim t^{-\alpha_B}$.
Using source average at $\Delta=0$, we measure effective exponents 
$\eta_C (t)$, $\eta_U (t)$, $\alpha_C (t)$, $\eta_B (t)$ and $1/z_U (t)$
up to $10^5$ Monte Carlo time steps. $\eta_C (t)$ is defined
as $\eta_C (t) = \log( N_C (mt)/ N_C (t))/ \log m$ 
and similarly for the others. 
Table shows the numerical results of source average.
$\eta_B$ is fairly underestimated due to the large correction with $\Delta_\eta = 0.08(1)$.
From Eq. (7) we predict $\alpha_C = 0.0725(100)$, $\alpha_U = 0.1525(125)$.
Because of $\eta_C > \eta_U$ and $\alpha_C < \alpha_U$, we predict
$\eta_B = \eta_C = 0.40(1)$ and $\alpha_B = \alpha_C = 0.0725(100)$. 
The prediction agree very with the numerical results.
Fig. 3 shows $\eta_C (t)$, $\eta_U (t)$ and $\eta_B (t)$. 
Our results are also equivalent to the previous result of the
second level of DP-DP coupling measured in slave average, 
$\eta_B = 0.39(2)$ and $\alpha_B = 0.075(10)$ \cite{uni}. 

Next, we consider a two-level hierarchy of systems belonging
to DP and PC class. 
We only discuss PC-DP coupling here. 

We consider the CP 
and branching annihilating random walks with two offsprings (BAW(2)) \cite{cp,baw2}.
BAW(2) belongs to PC class.
In BAW(2), a particle, $A$ hops to one of the $nn$ sites with probability 
$p$ and it creates two particles on two $nn$ sites to the left or right 
direction with $(1-\sigma)(1-p)$. 
If two particles happen to be on a same site by branching or hopping, 
they annihilate each other instantaneously.
The critical point of BAW(2) at $\sigma=0$ is at ${p_c}^A = 0.5105(7)$ \cite{baw2}.
The critical line is 
${\sigma_c}^{A} = 1- p R_A / (1-p)$, where $R_A = (1-{p_c}^A)/{p_c}^A = 0.95886$.
In CP, annihilation and branching of a $B$ particle occur with $p$ and $\sigma(1-p)$.
Criticality of CP at $\sigma=1$ is same as before. 
The critical line is now ${\sigma_c}^B = p R/ (1-p)$.
The dynamics $A \rightarrow A+B$ with $\sigma (1-p)$ couple BAW and CP without 
feedback from CP to BAW. An $A$ particle creates a $B$ 
particle on the same site of the CP level only if the target site is empty.   
The multicritical point $(\sigma_M , p_M)$ of BAW-CP coupling is 
$\sigma_M = 0.77474$ and $p_M = 0.19023$.

At the multicritical point, we estimate $\eta_C = 0.14(2)$, $\eta_U = 0.1875(25)$
and $2/z_U = 1.26(1)$. Using random initial conditions of 
$\rho_A (0) = \rho_B (0) = 1/2$, we also measure the exponent $\beta$ of ${\rho^s}_B$ in 
steady-states up to system size $1.2 \times 10^4$ and $\Delta = 6 \times 10^{-3}$. 
We estimate $\beta_B = 0.28(1)$. The numerical results show that the CP still belongs
to the DP class.

With $z_A = 1.750(5), \delta_A = 0.285(2)$ for BAW(2) \cite{baw4}, 
Eq. (7) predicts $\alpha_C = 0.146(20)$ and $\alpha_U = 0.1575(96)$. 
The value of $\alpha_C$ and $\alpha_U$ agree with the DP value 
$\alpha = 0.15964(6)$ \cite{hin} within numerical error. 
It means that $\rho_B$ uniformly decays but very slowly converges due to
the coupling effect in the coupled region.
Due to $\eta_C < \eta_U$, 
Eq. (7) predicts $\eta_B = \eta_U = 0.1875(25)$ which is not
the value of DP class, $\eta = 0.313686(8)$ \cite{hin}.
We directly measure $\eta_B = 0.18(1)$ which is somewhat underestimated but 
agrees well with the prediction. We conclude that
even though CP is not changed by BAW(2), Eq. (7) is satisfied very well
while Eq. (4) gives a wrong prediction for this coupling.
On the other hand, 
with the DP values for $\eta_B$ and $z_B$, Eq. (4) predicts $\alpha_B = 0.0338$,
which is completely different from the true DP value, $\alpha = 0.15964$.
Eq. (4) also gives a wrong prediction for BAW-CP-CP coupling in which BAW(2) 
is on the first level. The
CP of the third level is affected by the second level. 
But the second level is not affected by the first. 
In slave average, exponents of the third level 
are same as those of the second level in CP-CP coupling without BAW(2). 
These results cannot satisfy Eq. (4) because $\delta_1$ is not a DP value. 

In order to interprete these observations we stress that
the dynamic exponent $\eta$ can depend on average methods, but $\alpha$ does not
because it is the ratio of $\beta$ and $\nu_{||}$ which characterize off-critical
behavior. Since away from the critical point the survival probability for source
is unity and hence the source and the slave average coincide.
Therefore
the results of two average methods should give
the same value of $\alpha$. So it is natural that two values of $\alpha$ measured in
two different ways are same in CP-CP and BAW-CP coupling. 
The method of average is a matter of choice but one obtains distinct
scaling relations which are valid for any coupling.
It means that Eq. (4) is not a relation for the slave average because it is not
always satisfied by the results of slave average.  
Also Eq. (4) is not a general relation for unidirectionally coupled systems
because it is a relation for quantities of the total system. 
As the relation for quantities of a total system is obtained by comparing
Eq. (5) and (6), it can change according to the nature of the coupling.

In summary, unidirectionally coupled systems show an inhomogeneity in the number
of particles or density. A cluster in a higher hierarchy
is divided into a coupled and an uncoupled region.
In each region, dynamic quantities show different scaling behavior. 
In order to describe this property quantitatively we employ 
a special average method, the $source$ average. Using scaling arguments we derive 
two generalized scaling relations which describes the inhomogeneity. 
The presence of $\delta$ of the first level in the scaling relations arises
naturally from the source average. Our scaling relations also
explain the existence of large corrections to scaling of dynamic quantities of 
a total system which cause a long time drift of critical exponents. 
Our scaling relations should be tested for other couplings between different
universality classes and also in higher dimensions. The
DP-PC coupling will be a possible candidate for future study.

This work was supported by the Post-doctoral Fellowship Program of Korea
Science $\&$ Engineering Foundation (KOSEF).

\begin{table}
\caption{
Estimates of source average for the dynamic exponents of the CP-CP coupling. 
}
\begin{tabular}{ccccc}
\hline \hline
$\eta_C$  & $\eta_U$  &  $2/z_U$  &  $\alpha_C$  &  $\eta_B$ \\  
\hline
0.40(1)   & 0.32(1)   &  1.265(5) &  0.075(5)    &  0.375(25) \\
\hline \hline
\end{tabular}
\end{table}
\end{document}